\documentclass[aps,prd,reprint,nofootinbib,superscriptaddress,longbibliography]{revtex4-2}

\usepackage{amsmath,amssymb,bm,mathrsfs}
\usepackage{booktabs,array}
\usepackage[colorlinks=true,linkcolor=blue,citecolor=blue,urlcolor=blue]{hyperref}
\newcommand{\D}{\widetilde{\nabla}}
\newcommand{\curl}{\operatorname{curl}}

\newcommand{\tlR}{\widetilde R}

\newcommand{\la}{\langle}
\newcommand{\ra}{\rangle}
\newcommand{\TT}{\mathrm{TT}}

\begin{document}

\title{Constraint closure and gravitational-wave content of shear-free cosmologies in metric \texorpdfstring{$f(R)$}{f(R)} gravity}

\author{Mudhahir Al Ajmi}
\email{mudhahir@squ.edu.om}
\affiliation{Department of Physics, College of Science, Sultan Qaboos University,
Muscat 123, Oman}

\author{Amare Abebe}
\email{amare.abebe@nithecs.ac.za}
\affiliation{Center for Space Research, North-West University, Potchefstroom 2520, South Africa}
\affiliation{National Institute for Theoretical and Computational Sciences (NITheCS),
Potchefstroom 2520, South Africa}

\date{\today}

\begin{abstract}
We derive the consistency conditions governing the propagating content of
linear shear-free perturbations of
Friedmann-Lema\^{\i}tre-Robertson-Walker cosmologies in metric $f(R)$
gravity.  The divergence of the shear-free constraint is shown to be the total
momentum-conservation equation and therefore does not imply geodesic flow.
Its projected time derivative instead supplies a nontrivial integrability
condition.  Scalar, vector, and tensor sectors must then be tested separately.
Shear-freeness removes the independent transverse-traceless electric-magnetic
Weyl pair, so no ordinary $+$ or $\times$ tensor gravitational wave survives.
The scalaron is not removed algebraically, but its four-variable harmonic
system must remain in the time-dependent consistent subspace generated by the
shear-free constraint and all of its time derivatives.  For a geodesic
shear-free congruence in the spatially flat expanding de Sitter patch, this
excludes every nonzero fixed comoving scalar harmonic, including the healthy
$R+\alpha R^2-2\Lambda$ model with $\alpha,\Lambda>0$.  Vector modes obey a
curvature-dependent global eigenvalue condition whose right-hand side is
negative for positive-density matter with $f'>0$ and $1+w>0$; hence no
nonzero vector harmonic survives on that branch.  The familiar coasting
$R^3$ example evades this sign obstruction only because it lies on an
$f'<0$ branch.  Thus tensor radiation is absent in the imposed shear-free
sector, whereas scalar radiation is not excluded kinematically but remains a
model- and background-dependent constraint-closure question.
\end{abstract}

\maketitle

\section{Introduction}
\label{sec:introduction}

The kinematics of a relativistic matter congruence separates local volume
expansion, rotation, acceleration, and shape distortion.  Requiring the shear
$\sigma_{ab}$ to vanish is therefore a strong dynamical restriction rather
than a choice of coordinates.  For a barotropic perfect fluid close to a
Friedmann-Lema\^{\i}tre-Robertson-Walker (FLRW) background, the general
relativistic shear-free theorem excludes simultaneous expansion and rotation
under the usual physical assumptions \cite{Nzioki2011}.  Metric $f(R)$ gravity
was subsequently shown to modify the corresponding integrability condition,
and coasting power-law examples were presented as rotating-expanding
counterexamples \cite{AGD2011,AbebeConference2012}.  More recently, the role
of shear in supporting cosmological gravitational radiation was studied in
general relativity with imperfect matter \cite{Mayala2022,MayalaThesis2021}.

Three logically different questions occur in discussions concerning shear-free spacetimes:
\begin{itemize}
\item Does the shear-free constraint remain consistent under spatial and temporal
propagation? 
\item Does a formal second-order wave equation describe an
independent degree of freedom after all constraints are imposed? 
\item  Does the required local eigenvalue belong to the global harmonic spectrum of the
FLRW spatial sections?  
\end{itemize}
None of these questions can replace the other two.

The distinction is particularly important in metric $f(R)$ gravity.  Away
from the shear-free sector the theory contains the two tensor polarizations
and one massive scalar polarization \cite{Liang2017,Alves2024}.  A recent
kinematical analysis independently associates the transverse tensor response
with transverse shear, but the scalar response with expansion and the
longitudinal-breathing sector \cite{Maldonado2026}.  The scalar mode is not an
independent transverse-traceless (TT) Weyl tensor, and the elimination of the
TT pair does not by itself establish complete silence.  Conversely, the trace
equation alone does not establish that the scalaron is compatible with a
shear-free matter congruence.

This paper gives a model-by-model linear consistency procedure for metric $f(R)$
gravity under the following assumptions: (i) perturbations are taken about an
FLRW solution; (ii) $u^a$ is the congruence of minimally coupled barotropic
perfect-fluid matter, $p_m=w\mu_m$ with constant $w$, with $\mu_m>0$ and
$1+w>0$ wherever matter-normalized variables or matter momentum
conservation are used; (iii)
$\sigma_{ab}=0$ is imposed to first order and propagated along $u^a$; and
(iv) scalar, vector, and tensor modes are defined with respect to the
constant-curvature background spatial sections.  Conditions such as $F>0$ or
$G>0$, where $F=f'(R)$ and $G=f''(R)$, are imposed only when physical
viability is being tested.  The scalaron matrix additionally assumes
$G\ne0$ and $f\in C^4$.  The $G=0$ general-relativistic limit, which has no
scalaron, must be taken directly in the unsolved field and constraint
equations rather than in that matrix.  Exact vacuum is treated
separately because neither the matter-normalized density gradient nor a
matter-selected four-velocity exists there.

Related shear-free but homogeneous anisotropic $f(R)$ models
\cite{AbebeMomeni2016} and irrotational quasi-Newtonian scalar perturbations
\cite{SamiAbebe2020} impose different kinematic or matter assumptions and do
not supply the constraint closure derived here.  Likewise, the conformally
flat imperfect-fluid problem treated in Ref.~\cite{MayalaThesis2021} is not a
special case of $\sigma_{ab}=0$ unless shear-freeness is imposed separately.
Our use of the $1+3$ covariant equations and scalar perturbation variables
follows Refs.~\cite{EllisVanElst1999,Tsagas2008,Carloni2008}; a complementary
gauge-invariant metric treatment of $f(R)$ perturbations was given recently in
Ref.~\cite{Molano2025}.

Our principal results, and the organization of the manuscript, are as follows.
Section~\ref{sec:system} introduces the metric $f(R)$ system and the primary
shear-free constraint.  In Sec.~\ref{sec:consistency}, we establish that the
spatial divergence of this constraint reduces exactly to total momentum
conservation and therefore does not imply $A_a=0$, whereas its temporal
propagation produces a nontrivial projected symmetric trace-free (PSTF)
integrability condition.  Section~\ref{sec:content} resolves the resulting
constraints into their irreducible propagating sectors.  The tensor projection
removes the independent transverse-traceless Weyl pair; for nonvacuum matter,
the scalar sector closes as a four-dimensional linear system subject to a
recursively generated hierarchy of consistency conditions; and the vector
sector obeys a spatial spectral condition whose temporal propagation reproduces
the remaining rotation-expansion compatibility condition.  For $\mu_m>0$,
$1+w>0$, and $F>0$, the required vector eigenvalue has the wrong sign for every
regular constant-curvature vector harmonic.  In Sec.~\ref{sec:viability}, a
direct background audit shows explicitly that the proposed coasting $R^n$
examples evade this obstruction only on an unhealthy branch, and we examine
the corresponding constant-curvature vacuum constraints.  Section~\ref{sec:theorem}
collects these results into a linear shear-free classification theorem, while
Sec.~\ref{sec:conclusion} discusses their physical implications and conclusions.
Technical derivations and supporting identities are given in
Appendices~\ref{app:equations} - \ref{app:vector}.

\section{Metric \texorpdfstring{$f(R)$}{f(R)} system and shear-free constraint}
\label{sec:system}

We use units $8\pi G=c=1$ and signature $(-,+,+,+)$.  The generalized Einstein field equations
of metric $f(R)$ gravity may be written as
\begin{equation}
G_{ab}=\frac{T^m_{ab}}{F}+T^R_{ab}\equiv T_{ab},
\qquad F\equiv f'(R),
\label{eq:field}
\end{equation}
where the minimally coupled matter source is a barotropic perfect fluid,
\begin{equation}
T^m_{ab}=\mu_m u_a u_b+p_m h_{ab}\;,
\label{eq:matterT}
\end{equation}
with $\qquad p_m=w\mu_m,
\quad h_{ab}=g_{ab}+u_a u_b $ and
\begin{equation}
T^R_{ab}
=\frac{1}{F}\left[
\frac12(f-RF)g_{ab}
+\nabla_a\nabla_b F-g_{ab}\Box F
\right].
\label{eq:curvatureT}
\end{equation}
Relative to the matter four-velocity $u^a$, the total effective
energy-momentum tensor is decomposed as
\begin{equation}
T_{ab}
=\mu u_a u_b+p h_{ab}+2q_{(a}u_{b)}+\pi_{ab},
\label{eq:fluiddecomp}
\end{equation}
with
\begin{equation}
\mu=\frac{\mu_m}{F}+\mu_R,\qquad
p=\frac{p_m}{F}+p_R,\qquad
q_a=q^R_a,\qquad
\pi_{ab}=\pi^R_{ab},
\label{eq:totalfluid}
\end{equation}
where the last two equalities follow because the matter source is perfect.

At linear order about FLRW, in the shear-free sector, the effective curvature
fluid variables are
\begin{align}
\mu_R &=
\frac{1}{F}\left[
\frac12(RF-f)-\Theta G\dot R+G\widetilde{\nabla}^2R
\right],
\nonumber\\
p_R &=
\frac{1}{F}\left[
\frac12(f-RF)+G\ddot R+J\dot R^2
+\frac23\left(\Theta G\dot R-G\widetilde{\nabla}^2R\right)
\right],
\nonumber\\
q^R_a &=
-\frac{1}{F}\left[
J\dot R\,\widetilde{\nabla}_aR
+G\widetilde{\nabla}_a\dot R
-\frac13G\Theta\widetilde{\nabla}_aR
\right],
\nonumber\\
\pi^R_{ab} &=
\frac{G}{F}\widetilde{\nabla}_{\langle a}
\widetilde{\nabla}_{b\rangle}R,
\label{eq:curvfluid}
\end{align}
where $G\equiv f''(R)$ and $J\equiv f'''(R)$.  The total stress tensor is
conserved, while minimally coupled matter is also conserved separately.

The additional scalar degree of freedom is governed by the trace equation
\begin{equation}
3\Box F+RF-2f=T^m=3p_m-\mu_m,
\label{eq:trace}
\end{equation}
which, together with matter conservation, closes the scalar dynamics used
below.

All unlabelled thermodynamic variables below are total effective variables.
The total stress tensor is conserved, while minimally coupled matter is also
conserved separately.

At linear order the shear propagation equation is
\begin{equation}
\dot\sigma_{\la ab\ra}=-\frac23\Theta\sigma_{ab}
+\D_{\la a}A_{b\ra}-E_{ab}+\frac12\pi_{ab}.
\label{eq:shear-prop}
\end{equation}
Imposing $\sigma_{ab}=0$ promotes it to the primary constraint
\begin{equation}
(C_0)_{ab}\equiv\D_{\la a}A_{b\ra}-E_{ab}
+\frac12\pi_{ab}=0.
\label{eq:C0}
\end{equation}
The remaining Einstein-Bianchi constraints required below include:
\begin{align}
(C_1)_a&\equiv q_a-\frac23\D_a\Theta+(\curl\omega)_a=0,
\label{eq:C1}\\
(C_3)_{ab}&\equiv H_{ab}+\D_{\la a}\omega_{b\ra}=0,
\label{eq:C3}\\
(C_4)_a&\equiv\D^b\left(E_{ab}+\frac12\pi_{ab}\right)
-\frac13\D_a\mu+\frac13\Theta q_a=0.
\label{eq:C4}
\end{align}
Their full linear propagation is homogeneous; the genuinely new information
comes from preserving $(C_0)_{ab}$.

\section{Spatial and temporal consistency}
\label{sec:consistency}

\subsection{The spatial divergence is not an acceleration theorem}

Taking a divergence of Eq.~\eqref{eq:C0}, using the constant-curvature PSTF
identity, the Gauss relation, Eqs.~\eqref{eq:C1} and \eqref{eq:C4}, and the
vorticity propagation equation gives
\begin{equation}
\dot q_{\la a\ra}+\frac43\Theta q_a+\D_ap+\D^b\pi_{ab}
+(\mu+p)A_a=0.
\label{eq:momentum}
\end{equation}
This is precisely total momentum conservation.  No independent condition
$A_a=0$ follows.  Appendix~\ref{app:spatial} displays every intermediate
step, including the acceleration term in the scalar time-gradient
commutator,
\begin{equation}
(\D_af)^{\displaystyle\cdot}
=\D_a\dot f-\frac13\Theta\D_af+\dot f A_a.
\label{eq:scalar-comm}
\end{equation}

\subsection{Temporal propagation}

Writing Eq.~\eqref{eq:C0} as
$E_{ab}=\D_{\la a}A_{b\ra}+\pi_{ab}/2$, differentiating along $u^a$, and
using the electric-Weyl propagation equation together with
Eqs.~\eqref{eq:C1} and \eqref{eq:C3} gives
\begin{align}
0={}&\D_{\la a}\dot A_{b\ra}
+\frac23\Theta\D_{\la a}A_{b\ra}
+\frac13\D_{\la a}\D_{b\ra}\Theta
\nonumber\\
&+\dot\pi_{\la ab\ra}+\frac23\Theta\pi_{ab}.
\label{eq:temporal-general}
\end{align}
For metric $f(R)$ gravity, Eq.~\eqref{eq:curvfluid} and the full commutator reduce
this to
\begin{align}
0={}&\D_{\la a}\dot A_{b\ra}
+\left(\frac23\Theta+\frac{G\dot R}{F}\right)
\D_{\la a}A_{b\ra}
+\frac13\D_{\la a}\D_{b\ra}\Theta
\nonumber\\
&+\frac{G}{F}\D_{\la a}\D_{b\ra}\dot R
+\frac{FJ-G^2}{F^2}\dot R\,\D_{\la a}\D_{b\ra}R,
\label{eq:temporal-fR}
\end{align}
where $J=f'''(R)$.  For constant $w$, matter momentum conservation gives
$A_a=-w\D_a\mu_m/[(1+w)\mu_m]$ and hence
\begin{align}
0={}&\left(w+\frac13\right)
\left[\D_{\la a}\D_{b\ra}\Theta
+\Theta\D_{\la a}A_{b\ra}\right]
\nonumber\\
&+\frac{G}{F}\D_{\la a}\D_{b\ra}\dot R
+\frac{G\dot R}{F}\D_{\la a}A_{b\ra}
\nonumber\\
&+\frac{FJ-G^2}{F^2}\dot R\,
\D_{\la a}\D_{b\ra}R.
\label{eq:temporal-w}
\end{align}
Unlike the spatial divergence, Eq.~\eqref{eq:temporal-w} is a nontrivial
integrability condition.

\section{Irreducible propagating content}
\label{sec:content}

\subsection{Tensor sector: no ordinary gravitational waves}

The TT projection of Eq.~\eqref{eq:C0} removes scalar and vector distortions.
For perfect-fluid matter, Eq.~\eqref{eq:curvfluid} is a scalar PSTF Hessian and
has no TT part.  Therefore
\begin{equation}
E^{\TT}_{ab}=0.
\label{eq:E-TT}
\end{equation}
The TT projection of Eq.~\eqref{eq:C3} then yields
\begin{equation}
H^{\TT}_{ab}=0.
\label{eq:H-TT}
\end{equation}
The two tensors that form the Maxwell-like spin-2 propagation pair thus
vanish algebraically.  In the exactly shear-free sector there are no ordinary
$+$ and $\times$ tensor gravitational waves.  This statement is restricted
to the imposed matter congruence and does not remove the tensor modes of
general, non-shear-free $f(R)$ cosmologies.

\subsection{Scalar sector: invariant temporal closure}
\label{sec:scalar}

We first treat nonvacuum.  In this part $\mu_m>0$,
$1+w>0$, $G\ne0$, and $f\in C^4$; these hypotheses make the
matter-normalized gradient, acceleration, and scalaron evolution
matrix well defined.

Introduce the gauge-invariant scalar gradients
\begin{align}
\Delta_a&=\frac{a}{\mu_m}\D_a\mu_m,
&Z_a&=a\D_a\Theta,
\nonumber\\
{\cal R}_a&=a\D_aR,
&{\cal S}_a&=a\D_a\dot R,
\label{eq:gradients}
\end{align}
and scalar harmonics satisfying
$\D^2Q^{(k)}=-k^2Q^{(k)}/a^2$.  With
$\varkappa_k\equiv k^2/a^2$, define
\begin{equation}
{\cal U}_k=(\Delta_k,Z_k,{\cal R}_k,{\cal S}_k)^T.
\label{eq:U}
\end{equation}
Matter conservation, Raychaudhuri's equation, and the trace equation give a
closed first-order system
\begin{equation}
\dot{\cal U}_k={\mathsf M}_k{\cal U}_k,
\label{eq:matrix-system}
\end{equation}
whose full matrix is recorded in Appendix~\ref{app:scalar-matrix}.  Removing
the common nonzero scalar tensor harmonic from Eq.~\eqref{eq:temporal-w}
gives the shear-free scalar row
\begin{align}
{\mathscr C}_k={}&\left(w+\frac13\right)
\left(Z_k-\frac{w\Theta}{1+w}\Delta_k\right)
\nonumber\\
&+\frac{G}{F}\left({\cal S}_k
-\frac{w\dot R}{1+w}\Delta_k\right)
+\frac{FJ-G^2}{F^2}\dot R\,{\cal R}_k=0.
\label{eq:scalar-row}
\end{align}
Writing ${\mathscr C}_k={\mathsf L}_0{\cal U}_k$, successive temporal
consistency rows are generated without an arbitrary truncation by
\begin{equation}
{\mathsf L}_{r+1}=\dot{\mathsf L}_r+{\mathsf L}_r{\mathsf M}_k,
\qquad {\mathsf L}_r{\cal U}_k=0,
\qquad r=0,1,2,\ldots .
\label{eq:row-recursion}
\end{equation}
At each time define the consistent subspace
$\mathcal K_k(t)=\bigcap_{r\ge0}\ker {\mathsf L}_r(t)$.  A constrained
solution must remain in $\mathcal K_k(t)$ throughout the interval; equivalently,
its trajectory must be tangent to this time-dependent subspace under
Eq.~\eqref{eq:matrix-system}.  If the background, $f$, and the perturbation
coefficients are analytic, imposing all rows at one initial time is locally
sufficient.  Without analyticity, Eq.~\eqref{eq:row-recursion} is an interval
condition rather than a claim that an infinite vanishing jet at one time is
sufficient.  A wave equation for $\delta R_k$ alone is therefore necessary but
not sufficient, and the existence of a nonzero scalar mode remains a
model-by-model question.

Exact vacuum is not obtained by setting $\mu_m=0$ in
Eq.~\eqref{eq:gradients}, since $\Delta_a$ and the matter-selected congruence
then cease to be defined.  We therefore consider a separate, explicitly
restricted vacuum problem: the standard geodesic congruence, $A_a=0$ to first
order, in the spatially flat expanding de Sitter patch.  Its background data
are
\begin{equation}
R_0F_0=2f_0,
\qquad \dot R_0=0,
\qquad \tlR_0=0,
\qquad R_0=\frac43\Theta_0^2,
\label{eq:dS-data}
\end{equation}
where $\Theta_0>0$ is constant.  The last equality follows in this flat
slicing; it is not an assertion about the closed or open FLRW slicings of de
Sitter spacetime.

For the reduced vacuum state $({Z}_k,{\cal R}_k,{\cal S}_k)$,
Eqs.~\eqref{eq:temporal-fR}, \eqref{eq:Friedmann}, and
\eqref{eq:trace-bg} give
\begin{align*}
\dot Z_k={}&-\frac23\Theta_0 Z_k
+\left(\frac12-\frac{R_0G_0}{4F_0}
+\frac{G_0}{F_0}\varkappa_k\right){\cal R}_k
\nonumber\\
&+\frac{\Theta_0G_0}{F_0}{\cal S}_k,\\
\dot{\cal R}_k={}&{\cal S}_k,\\
\dot{\cal S}_k={}&-\left(\varkappa_k+\frac{F_0}{3G_0}
-\frac{R_0}{3}\right){\cal R}_k-\Theta_0{\cal S}_k.
\end{align*}
The scalar projection of Eq.~\eqref{eq:temporal-fR} is
\begin{equation} 
Z_k/3+(G_0/F_0){\cal S}_k=0\;.
\end{equation}
  Differentiating this unevaluated relation,
using the three equations above, and only then eliminating ${\cal S}_k$ gives
\begin{equation*}
0=\left[-\frac16+\frac{R_0G_0}{4F_0}
-\frac{2G_0}{3F_0}\varkappa_k\right]{\cal R}_k.
\end{equation*}
Thus, for $G_0\ne0$ and a nonzero curvature amplitude,
\begin{equation}
\varkappa_k=\frac{3R_0}{8}-\frac{F_0}{4G_0}.
\label{eq:dS-k}
\end{equation}
The right-hand side is constant, while
$\dot\varkappa_k=-2\Theta_0\varkappa_k/3$.  Hence no fixed nonzero comoving
scalar harmonic can satisfy Eq.~\eqref{eq:dS-k} throughout an expanding de
Sitter interval.  The homogeneous $k=0$ mode is exceptional because its PSTF
Hessian vanishes.  A nongeodesic vacuum congruence would introduce an
independent acceleration potential and is not covered by this no-go result.

\subsection{Vector sector: local, temporal, and global tests}
\label{sec:vector}

Let $K=0,+1,-1$ and $\tlR=6K/a^2$.  Expand the transverse vorticity in vector
harmonics,
\begin{equation}
\omega_a=\sum_\nu\omega_\nu Q_a^{(\nu)},
\quad \D^aQ_a^{(\nu)}=0,
\quad \D^2Q_a^{(\nu)}=-\frac{k_V^2}{a^2}Q_a^{(\nu)}.
\label{eq:vector-harmonics}
\end{equation}
For the simply connected covers $\mathbb R^3$, $S^3$, and $H^3$, the standard
constant-curvature spectrum is \cite{KodamaSasaki1984}
\begin{center}
\begin{tabular}{ccc}
\toprule
$K$ & $k_V^2$ & $k_V^2-2K$\\
\midrule
$0$ & $\nu^2$ & $\nu^2$\\
$+1$ & $\nu^2-2$, $\nu=2,3,\ldots$ & $\nu^2-4$\\
$-1$ & $\nu^2+2$ & $\nu^2+4$\\
\bottomrule
\end{tabular}
\end{center}
Here $\nu\ge0$ is continuous for $K=0,-1$.  Compact flat or hyperbolic
quotients have topology-dependent discrete spectra; in that case the table
must be replaced by the spectrum of the chosen quotient, while the local
compatibility equation below is unchanged.
There are two equivalent routes to the complete vector compatibility test.
First, take the curl of Eq.~\eqref{eq:C1}, set $q_a=q^R_a$, and use
Eqs.~\eqref{eq:curvfluid}, \eqref{eq:curlcurl-app}, and the background
Raychaudhuri and Gauss equations.  The result is
\begin{align}
\left[\frac{k_V^2-2K}{a^2}+2N\right]\omega_\nu&=0,
&N&=\frac{(1+w)\mu_m}{F}.
\label{eq:vector-spectrum}
\end{align}
For a nonzero mode the instantaneous physical eigenvalue must therefore
satisfy
\begin{equation}
\frac{k_V^2-2K}{a^2}=-2N.
\label{eq:vector-eigenvalue}
\end{equation}
Because $k_V^2-2K\ge0$ on each spectrum in the table, Eq.~\eqref{eq:vector-eigenvalue}
has no nonzero solution when $\mu_m>0$, $1+w>0$, and $F>0$.  The $K=+1$,
$\nu=2$ Killing mode saturates the left-hand side but would require $N=0$.

For completeness, differentiating the unevaluated local relation, using
$\dot\omega_{\la a\ra}=(w-2/3)\Theta\omega_a$ and the vector Laplacian
commutator, gives the remaining compatibility condition
\begin{equation}
N\left(C+\frac13\Theta\right)\omega_a=0,
\qquad C=w\Theta+\frac{\dot F}{F}.
\label{eq:vector-temporal}
\end{equation}
The same equation follows without a time derivative by taking the curl of
the $(C_0)$ - $(C_4)$ compatibility vector and then using
Eq.~\eqref{eq:vector-local-app}.  Appendix~\ref{app:vector} gives both
derivations without division by $C$, $N$, or $\Theta$.  Thus a local
d'Alembertian equation is not a mode-existence criterion: the spatial
constraint, its temporal closure, and the global spectrum must all agree.

\section{Viability and background audits}
\label{sec:viability}

The compatibility relations have physical content only on a background that
solves
\begin{align}
\frac{\Theta^2}{3}+\frac{\tlR}{2}
&=\frac1F\left[\mu_m+\frac12(RF-f)-\Theta\dot F\right],
\label{eq:Friedmann}\\
G\ddot R+J\dot R^2+\Theta G\dot R
&=\frac13\left[RF-2f+(1-3w)\mu_m\right],
\label{eq:trace-bg}\\
\dot\mu_m&=-(1+w)\Theta\mu_m.
\label{eq:matter-bg}
\end{align}

\subsection{No healthy coasting \texorpdfstring{$R^n$}{R-to-the-n} vector branch}

For $f(R)=\beta R^n$ on a flat power-law background $a\propto t^p$,
matching the time powers gives
\begin{equation}
p=\frac{2n}{3(1+w)}.
\label{eq:p-n}
\end{equation}
The coasting case $p=1$ therefore requires $3(1+w)=2n$.  With the convenient
normalization $a=R^{-1/2}$, so $R=6/t^2$ and
$\mu_m=\mu_0R^n$, direct substitution in Eqs.~\eqref{eq:Friedmann} and
\eqref{eq:trace-bg} gives
\begin{equation}
\frac{\mu_0}{\beta}=\frac{1-2n(n-1)}2.
\label{eq:amplitude}
\end{equation}
The numerical value assigned to the comoving label $k$ changes under a
constant rescaling of $a$, but $k^2/a^2$ and the sign test below do not.
The temporal vector condition is satisfied because $C=-\Theta/3$, while the
spatial condition requires
\begin{equation}
k^2=\frac{4n(n-1)-2}{3}.
\label{eq:k-n}
\end{equation}
Let $A_n=1-2n(n-1)$.  Then
\begin{equation}
\frac{\mu_0}{\beta}=\frac{A_n}{2},
\qquad k^2=-\frac{2A_n}{3}.
\label{eq:sign-pair}
\end{equation}
For $\mu_0>0$ and $k^2>0$, one must have $A_n<0$ and hence $\beta<0$.
Since $R>0$,
\begin{equation}
F=n\beta R^{n-1},
\qquad G=n(n-1)\beta R^{n-2}.
\label{eq:FG-power}
\end{equation}
The condition $A_n<0$ restricts $n$ to
$n>(1+\sqrt3)/2$ or $n<(1-\sqrt3)/2$, so $n=0$ is not a candidate.  On the
positive-$n$ branch, $F<0$; on the negative-$n$ branch, $F>0$ but $G<0$.
Consequently no real $n$ supports positive density, a nonzero regular
coasting vector harmonic, and both viability signs $F>0$, $G>0$.  The
$n=3$, $w=1$ example has
\begin{equation}
\beta=-\frac{2\mu_0}{11},
\qquad k^2=\frac{22}{3},
\qquad F=3\beta R^2<0.
\label{eq:R3}
\end{equation}
Its eigenvalue belongs to the continuous vector spectrum of the simply
connected flat section, but its effective gravitational coupling has the
wrong sign.

\subsection{Stable constant-curvature vacua}

At a viable de Sitter point one normally requires \cite{DeFelice2010}
\begin{equation}
F_0>0,
\qquad G_0>0,
\qquad
m_s^2=\frac{F_0}{3G_0}-\frac{R_0}{3}\ge0.
\label{eq:health}
\end{equation}
Equation~\eqref{eq:dS-k} can then be written
\begin{equation}
\varkappa_k=\frac{R_0}{8}-\frac{3m_s^2}{4}.
\label{eq:dS-health-k}
\end{equation}
Irrespective of the sign of its constant right-hand side, the redshifting
$\varkappa_k$ cannot remain equal to it for a fixed nonzero $k$ during
expansion.  Within the geodesic shear-free congruence of the spatially flat
expanding patch specified below Eq.~\eqref{eq:dS-data}, this gives a no-go
result for the healthy constant-curvature class, not only for a selected
Lagrangian.

As an explicit model, take the quadratic Starobinsky-type Lagrangian with a
cosmological constant
\cite{Starobinsky1980}
\begin{equation}
f(R)=R+\alpha R^2-2\Lambda,
\qquad \alpha>0,
\quad \Lambda>0.
\label{eq:starobinsky-Lambda}
\end{equation}
The vacuum solution and stability data are
\begin{equation}
R_0=4\Lambda,
\quad F_0=1+8\alpha\Lambda,
\quad G_0=2\alpha,
\quad m_s^2=\frac{1}{6\alpha},
\label{eq:starobinsky-data}
\end{equation}
so the assumed shear-free scalar wave number is
\begin{equation}
\varkappa_k=\frac{\Lambda}{2}-\frac{1}{8\alpha}.
\label{eq:starobinsky-k}
\end{equation}
Even when this is positive, equality holds at most instantaneously.  The
geodesic shear-free perturbation sector of this flat expanding de Sitter patch
therefore contains neither the tensor pair nor a nonzero fixed comoving scalar
harmonic.

\section{Classification theorem}
\label{sec:theorem}

\noindent\textit{Linear shear-free classification.}\par
\noindent Consider linear perturbations of a metric $f(R)$ FLRW solution with
$F\ne0$, resolved relative to a timelike congruence on which
$\sigma_{ab}=0$ is imposed and propagated.  For minimally coupled perfect-fluid matter the following statements hold:
\begin{enumerate}
\item The TT projections obey $E^{\TT}_{ab}=H^{\TT}_{ab}=0$, so no independent
spin-2 Weyl mode exists in the shear-free sector.
\item For $p_m=w\mu_m$ with constant $w$, $\mu_m>0$, $1+w>0$, $G\ne0$, and
$f\in C^4$, a scalar solution exists only if its trajectory remains in the
time-dependent consistent subspace generated by
Eq.~\eqref{eq:row-recursion}.
\item Under the same fluid assumptions, except that no condition on $G$ is
needed, a nonzero vector mode must satisfy Eqs.~\eqref{eq:vector-eigenvalue}
and \eqref{eq:vector-temporal} for an allowed global harmonic.  In particular,
$F>0$ excludes every such mode.
\item In exact vacuum the matter-normalized formulation does not apply.  For
the separately specified geodesic congruence in the spatially flat expanding
de Sitter patch, Eq.~\eqref{eq:dS-k} excludes each nonzero fixed comoving
scalar harmonic.
\end{enumerate}
A second-order wave equation by itself proves none of these mode-existence
statements.  The $G=0$ general-relativistic limit must be evaluated directly
in the original field and constraint equations, before any division by $G$.

This theorem is deliberately restricted to linear perturbations about FLRW,
metric $f(R)$ gravity, a minimally coupled barotropic perfect fluid, and the
condition $\sigma_{ab}=0$.  It is not a theorem about nonlinear waves,
shear-free null congruences, anisotropic background cosmologies, or general
imperfect matter.  Nor does the vacuum statement cover an arbitrarily chosen
nongeodesic congruence.

\section{Discussion and conclusions}
\label{sec:conclusion}

The main result  of our study can be stated as follows:  in the exactly shear-free spacetimes, no ordinary tensor
gravitational waves propagate as the 
$+$ and $\times$ Weyl pair vanishes.  For scalar gravitational radiation
shear-freeness alone is not sufficient to make a universal statement.  The
scalaron survives kinematically, but an admissible mode must obey both its
dynamical equations and the full temporal closure of the shear-free
constraint.  In the geodesic, spatially flat, expanding de Sitter vacuum
problem treated here, that closure excludes every nonzero fixed comoving
scalar harmonic, including the healthy quadratic model in
Eq.~\eqref{eq:starobinsky-Lambda}.

This conclusion refines rather than repeats earlier literature.  The GR
silence result of Ref.~\cite{Mayala2022} does not transfer unchanged to the
effective curvature fluid because the spatial divergence of the shear-free
constraint is total momentum conservation and does not set $A_a$ to zero.
The rotation-expansion relation of Ref.~\cite{AGD2011} is one part of the
constraint system.  Combining it with the heat-flux form of
Eq.~\eqref{eq:C1}, or equivalently propagating the latter relation in time,
gives Eq.~\eqref{eq:vector-temporal}.  More decisively,
Eq.~\eqref{eq:vector-eigenvalue} has the wrong sign for every regular
constant-curvature vector harmonic when $\mu_m>0$, $1+w>0$, and $F>0$.
The coasting $\beta R^n$ family illustrates the complementary unhealthy
case: a regular nonzero harmonic requires either $F<0$ or $G<0$.

General metric $f(R)$ gravity outside
the imposed $\sigma_{ab}=0$ sector still possesses two tensor modes and a
scalaron.  Here the tensor pair is removed by the shear-free constraints,
whereas scalar radiation is decided dynamically by constraint closure.  This
is consistent with the independent kinematical identification of transverse
tensor polarization with transverse shear and of the scalar polarization with
expansion and longitudinal-breathing response \cite{Maldonado2026}.  A formal
d'Alembertian equation for vorticity or curvature is not sufficient evidence
for a propagating mode.

The natural extension is to general imperfect fluid sources with causal
transport.  Such a calculation must distinguish matter heat flux and
anisotropic stress from the effective curvature variables in
Eqs.~ \eqref{eq:curvfluid}; imposing a transport law on the
total effective flux would mix dynamics with a geometric rearrangement of the
field equations.

\begin{acknowledgments}
AA acknowledges the hospitality of the Department of Physics at Sultan Qaboos
University during his research visit, during which most of this work was completed.
M Al-Ajmi (ORCID ID 0000-0001-9888-5318) thanks Sultan Qaboos University for support with grants IG/--/SCI/P/25/77.
AI tools were used under the
authors' direction to assist with prose reorganization and independent algebraic and symbolic checks.  The authors supplied the
physical assumptions,  independently
rederived and verified every reported equation and claim, revised all
AI-assisted output, and take full responsibility for the content.
\end{acknowledgments}

\section*{Data availability}
No data were created or analyzed in this theoretical study.

\appendix

\section{Linearized equations and identities}
\label{app:equations}

The propagation equations used in the main text are:
\begin{align}
&\dot\Theta-\D^aA_a=-\frac13\Theta^2-\frac12(\mu+3p),
\label{eq:ray-app}\\
&\dot\omega_{\la a\ra}-\frac12(\curl A)_a
=-\frac23\Theta\omega_a,
\label{eq:omega-app}\\
&\dot E_{\la ab\ra}=(\curl H)_{ab}-\Theta E_{ab}
-\frac16\Theta\pi_{ab}
\nonumber\\
&\quad-\frac12\dot\pi_{\la ab\ra}
-\frac12\D_{\la a}q_{b\ra},
\label{eq:E-app}\\
&\dot H_{\la ab\ra}=-(\curl E)_{ab}
+\frac12(\curl\pi)_{ab}-\Theta H_{ab}.
\label{eq:H-app}
\end{align}
The constant-curvature spatial identities required for a first-order vector
$V_a$ and PSTF tensor $S_{ab}$ are:
\begin{align}
&\D^b\D_{\la a}V_{b\ra}
=\frac12\D^2V_a+\frac16\D_a(\D^bV_b)+\frac16\tlR V_a,
\label{eq:divdist-app}\\
&(\curl\curl V)_a
=\D_a(\D^bV_b)-\D^2V_a+\frac13\tlR V_a,
\label{eq:curlcurl-app}\\
&\curl(\D_{\la a}V_{b\ra})
=\frac12\D_{\la a}(\curl V)_{b\ra},
\label{eq:curlpstf-app}\\
&\curl(\D f)_a=2\dot f\,\omega_a.
\label{eq:curlgrad-app}
\end{align}
For a first-order spatial vector on FLRW geometry,
\begin{equation}
(\D^2V_a)^{\displaystyle\cdot}
=\D^2\dot V_{\la a\ra}-\frac23\Theta\D^2V_a.
\label{eq:lap-comm-app}
\end{equation}
For an explicit check, use comoving FLRW coordinates
$h_{ij}=a^2(t)\gamma_{ij}$, where the connection of $\gamma_{ij}$ is
time-independent.  If $\Delta_\gamma$ is the corresponding covector
Laplacian and $H=\dot a/a=\Theta/3$, then
\begin{equation*}
\D^2V_i=a^{-2}\Delta_\gamma V_i,
\qquad
\dot V_{\la i\ra}=\partial_tV_i-HV_i.
\end{equation*}
It follows directly that
\begin{align*}
(\D^2V_i)^{\displaystyle\cdot}
&=a^{-2}\Delta_\gamma(\partial_tV_i)-3H\D^2V_i,\\
\D^2\dot V_{\la i\ra}
&=a^{-2}\Delta_\gamma(\partial_tV_i)-H\D^2V_i,
\end{align*}
whose difference proves Eq.~\eqref{eq:lap-comm-app}.  Spatial curvature
enters when spatial derivatives are reordered, as in
Eq.~\eqref{eq:curlcurl-app}, but it does not generate an additional term in
this time-Laplacian commutator. 

\section{Spatial divergence of the shear-free constraint}
\label{app:spatial}

From Eq.~\eqref{eq:C0},
\begin{equation}
\D^bE_{ab}=\D^b\D_{\la a}A_{b\ra}+\frac12\D^b\pi_{ab}.
\label{eq:spatial-1}
\end{equation}
Using Eqs.~\eqref{eq:divdist-app} and \eqref{eq:curlcurl-app},
\begin{equation}
\D^b\D_{\la a}A_{b\ra}
=\frac23\D_a(\D^bA_b)+\frac13\tlR A_a
-\frac12(\curl\curl A)_a.
\label{eq:spatial-2}
\end{equation}
The background Gauss-Codazzi relation is expressed as
$\tlR=2(\mu-\Theta^2/3)$.  Curling Eq.~\eqref{eq:omega-app} and using
Eq.~\eqref{eq:C1}, including Eq.~\eqref{eq:scalar-comm} for
$\D_a\Theta$, gives
\begin{align}
\frac12(\curl\curl A)_a={}&\frac23\D_a(\D^bA_b)
-\frac13\D_a(\mu+3p)-\dot q_{\la a\ra}
\nonumber\\
&-\Theta q_a+\frac23\dot\Theta A_a.
\label{eq:spatial-3}
\end{align}
Substitution in Eq.~\eqref{eq:spatial-2}, followed by Eq.~\eqref{eq:C4},
gives
\begin{align}
0={}&\left[\frac23\left(\mu-\frac13\Theta^2\right)
-\frac23\dot\Theta\right]A_a
+\D_ap+\dot q_{\la a\ra}
\nonumber\\
&+\frac43\Theta q_a+\D^b\pi_{ab}.
\label{eq:spatial-4}
\end{align}
The background Raychaudhuri equation changes the square bracket in
Eq.~\eqref{eq:spatial-4} to $\mu+p$, proving Eq.~\eqref{eq:momentum}.
Omitting the acceleration term from Eq.~\eqref{eq:scalar-comm} prevents this
reduction and creates a spurious acceleration constraint.

\section{Temporal derivative and scalar matrix}
\label{app:scalar-matrix}

Throughout this appendix $\mu_m>0$, $1+w>0$, $G\ne0$, and $f\in C^4$.
These are domain conditions for the chosen variables and for divisions by
$1+w$ and $G$, not extra stability assumptions.

Differentiating Eq.~\eqref{eq:C0} gives
\begin{align}
\dot E_{\la ab\ra}={}&\D_{\la a}\dot A_{b\ra}
-\frac13\Theta\D_{\la a}A_{b\ra}
+\frac12\dot\pi_{\la ab\ra}.
\label{eq:time-1}
\end{align}
Equating this with Eq.~\eqref{eq:E-app} and eliminating $E_{ab}$ with
Eq.~\eqref{eq:C0} gives
\begin{align}
0={}&\D_{\la a}\dot A_{b\ra}
+\frac23\Theta\D_{\la a}A_{b\ra}
+\dot\pi_{\la ab\ra}+\frac23\Theta\pi_{ab}
\nonumber\\
&+\frac12\D_{\la a}q_{b\ra}-(\curl H)_{ab}.
\label{eq:time-2}
\end{align}
Equations~\eqref{eq:C1}, \eqref{eq:C3}, and
\eqref{eq:curlpstf-app} imply
\begin{equation}
(\curl H)_{ab}=-\frac13\D_{\la a}\D_{b\ra}\Theta
+\frac12\D_{\la a}q_{b\ra}.
\label{eq:time-3}
\end{equation}
The heat-flux terms cancel, yielding Eq.~\eqref{eq:temporal-general}.

For completeness, set $P=f^{(4)}$ and define
\begin{align}
{\cal V}&=\frac{RF-2f+(1-3w)\mu_m}{3G},
\label{eq:Vdef}\\
{\cal V}_{,R}&=\frac{RG-F}{3G}
-\frac{J[RF-2f+(1-3w)\mu_m]}{3G^2}.
\label{eq:VRdef}
\end{align}
The four scalar equations are
\begin{align}
\dot\Delta_k&=w\Theta\Delta_k-(1+w)Z_k,
\label{eq:Ddot-app}\\
\dot Z_k&=a_\Delta\Delta_k+a_ZZ_k+a_R{\cal R}_k+a_S{\cal S}_k,
\label{eq:Zdot-app}\\
\dot{\cal R}_k&={\cal S}_k-\frac{w\dot R}{1+w}\Delta_k,
\label{eq:Rdot-app}\\
\dot{\cal S}_k&=b_\Delta\Delta_k-\dot R Z_k
+b_R{\cal R}_k+b_S{\cal S}_k,
\label{eq:Sdot-app}
\end{align}
where
\begin{align}
a_\Delta&=-\frac{\mu_m}{F}
+\frac{w}{1+w}(\varkappa_k-\dot\Theta),
\label{eq:aDelta-coeff}\\
a_Z&=-\frac23\Theta+\frac{G\dot R}{F},
\label{eq:aZ-coeff}\\
a_S&=\frac{\Theta G}{F},
\label{eq:a-coeff-short}\\
a_R&=\frac{\mu_mG}{F^2}+\frac{F^2-fG}{2F^2}
+\frac{\Theta\dot R(JF-G^2)}{F^2}
\nonumber\\
&\quad+\frac{G}{F}\varkappa_k,
\label{eq:a-coeff}\\
b_\Delta&=\frac{(1-3w)\mu_m}{3G}
-\frac{w}{1+w}\ddot R,
\label{eq:bDelta-coeff}\\
b_S&=-\Theta-2\frac{J}{G}\dot R,
\label{eq:bS-coeff}\\
b_R&={\cal V}_{,R}
-\frac{P}{G}\dot R^2+\frac{J^2}{G^2}\dot R^2-\varkappa_k.
\label{eq:b-coeff}
\end{align}
Thus
\begin{equation}
{\mathsf M}_k=
\begin{pmatrix}
w\Theta&-(1+w)&0&0\\
a_\Delta&a_Z&a_R&a_S\\
-w\dot R/(1+w)&0&0&1\\
b_\Delta&-\dot R&b_R&b_S
\end{pmatrix},
\label{eq:M-app}
\end{equation}
which completes the constructive definition of the recursion in
Eq.~\eqref{eq:row-recursion}.

\section{Vector propagation and background signs}
\label{app:vector}

We first derive the local condition used in Sec.~\ref{sec:vector}.  Taking
the curl of Eq.~\eqref{eq:C1}, using $\D^a\omega_a=0$ and
Eq.~\eqref{eq:curlcurl-app}, gives
\begin{equation*}
(\curl q)_a=\frac43\dot\Theta\,\omega_a
+\D^2\omega_a-\frac13\tlR\omega_a.
\end{equation*}
On the other hand, Eqs.~\eqref{eq:curvfluid} and
\eqref{eq:curlgrad-app} give, for perfect fluid,
\begin{equation*}
(\curl q^R)_a=-\frac{2}{F}
\left(\ddot F-\frac13\Theta\dot F\right)\omega_a.
\end{equation*}
The two background identities needed to equate these expressions are
\begin{align*}
\mu+p&=N+\frac{1}{F}
\left(\ddot F-\frac13\Theta\dot F\right)
=\frac13\tlR-\frac23\dot\Theta,\\
N&=\frac{(1+w)\mu_m}{F}.
\end{align*}
Substitution cancels $\dot\Theta$, $\ddot F$, and $\Theta\dot F$ explicitly
and leaves
\begin{equation}
\D^2\omega_a+\left(\frac13\tlR-2N\right)\omega_a=0.
\label{eq:vector-local-app}
\end{equation}
This is Eq.~\eqref{eq:vector-spectrum} before harmonic decomposition.

To propagate it, define
$\lambda=\tlR/3-2N$ and
${\cal S}_a=\D^2\omega_a+\lambda\omega_a$.  Using
$\dot\omega_{\la a\ra}=(w-2/3)\Theta\omega_a$ and
Eq.~\eqref{eq:lap-comm-app}, its unevaluated time derivative is
\begin{align}
0=\dot{\cal S}_{\la a\ra}={}&
\left(w-\frac43\right)\Theta\D^2\omega_a
\nonumber\\
&+\left[\dot\lambda
+\left(w-\frac23\right)\Theta\lambda\right]\omega_a.
\label{eq:vector-undivided}
\end{align}
Eliminating $\D^2\omega_a=-\lambda\omega_a$ only after differentiation gives
\begin{align*}
0&=\left(\dot\lambda+\frac23\Theta\lambda\right)\omega_a\\
&=2N\left(C+\frac13\Theta\right)\omega_a,
\qquad
C=w\Theta+\frac{\dot F}{F},
\end{align*}
where
$\dot{\tlR}=-2\Theta\tlR/3$ and
$\dot N=-(\Theta+C)N$ were used.  This proves
Eq.~\eqref{eq:vector-temporal} without division by any background quantity.

As an independent check, curling the $(C_0)$ - $(C_4)$ compatibility vector
as in Ref.~\cite{AGD2011} gives
\begin{equation*}
C\left(\D^2\omega_a+\frac13\tlR\omega_a\right)
+\frac23\Theta N\omega_a=0.
\end{equation*}
Equation~\eqref{eq:vector-local-app} turns the first bracket into
$2N\omega_a$, and the same Eq.~\eqref{eq:vector-temporal} follows.  Thus the
published rotation-expansion relation is a necessary part of the constraint
system but is not, by itself, the complete spectral test.  In exact
vacuum the congruence must first be specified independently, as in the scalar
discussion; the matter-derived definitions of $C$ and the vorticity
propagation law cannot simply be continued through $\mu_m=0$.

For the coasting $\beta R^n$ background, the Friedmann equation reads
\begin{equation}
\frac12R=\frac{\mu_0}{n\beta}R
+\frac{n-1}{2n}R+(n-1)R,
\label{eq:coast-Friedmann-app}
\end{equation}
which immediately gives Eq.~\eqref{eq:amplitude}.  The trace equation gives
\begin{align}
G\ddot R+J\dot R^2+\Theta G\dot R
&=\frac23n(n-1)(n-2)\beta R^n,
\nonumber\\
\frac13[RF-2f+(1-3w)\mu_m]
&=\frac{n-2}{3}(\beta-2\mu_0)R^n.
\label{eq:coast-trace-app}
\end{align}
For $n\ne2$, equality reproduces Eq.~\eqref{eq:amplitude}.  At $n=2$ the
trace equation degenerates to $0=0$, but the Friedmann equation still fixes
the amplitude.  Hence the background check is not based on a freely chosen
normalization.  After Eq.~\eqref{eq:amplitude} is imposed, both sides above
reduce to
\begin{equation*}
\frac23n(n-1)(n-2)\beta R^n.
\end{equation*}


\begin{thebibliography}{99}
\raggedright

\bibitem{Nzioki2011}
A.~M. Nzioki, R. Goswami, P.~K.~S. Dunsby, and G.~F.~R. Ellis,
\emph{Shear-free perturbations of
Friedmann-Lema\^{\i}tre-Robertson-Walker universes},
Phys. Rev. D \textbf{84}, 124028 (2011),
\href{https://doi.org/10.1103/PhysRevD.84.124028}{doi:10.1103/PhysRevD.84.124028}.

\bibitem{AGD2011}
A. Abebe, R. Goswami, and P.~K.~S. Dunsby,
\emph{Shear-free perturbations of $f(R)$ gravity},
Phys. Rev. D \textbf{84}, 124027 (2011),
\href{https://doi.org/10.1103/PhysRevD.84.124027}{doi:10.1103/PhysRevD.84.124027}.

\bibitem{AbebeConference2012}
A. Abebe, R. Goswami, and P.~K.~S. Dunsby,
\emph{Simultaneous expansion and rotation of shear-free universes in
modified gravity}, AIP Conf. Proc. \textbf{1458}, 307 - 310 (2012),
\href{https://doi.org/10.1063/1.4734421}{doi:10.1063/1.4734421}.

\bibitem{Mayala2022}
R.~M. Mayala, R. Goswami, and S.~D. Maharaj,
\emph{Role of spacetime shear for cosmological gravitational waves, in
presence of imperfect fluids},
Int. J. Mod. Phys. D \textbf{31}, 2250105 (2022),
\href{https://doi.org/10.1142/S021827182250105X}{doi:10.1142/S021827182250105X}.

\bibitem{MayalaThesis2021}
R.~M. Mayala, \emph{Role of Weyl tensor and spacetime shear in relativistic
fluids}, Ph.D. thesis, University of KwaZulu-Natal (2021),
\href{https://researchspace.ukzn.ac.za/handle/10413/20213}{ResearchSpace UKZN}.

\bibitem{Liang2017}
D. Liang, Y. Gong, S. Hou, and Y. Liu,
\emph{Polarizations of gravitational waves in $f(R)$ gravity},
Phys. Rev. D \textbf{95}, 104034 (2017),
\href{https://doi.org/10.1103/PhysRevD.95.104034}{doi:10.1103/PhysRevD.95.104034}.

\bibitem{Alves2024}
M.~E.~S. Alves,
\emph{Testing gravity with gauge-invariant polarization states of
gravitational waves: Theory and pulsar timing sensitivity},
Phys. Rev. D \textbf{109}, 104054 (2024),
\href{https://doi.org/10.1103/PhysRevD.109.104054}{doi:10.1103/PhysRevD.109.104054}.

\bibitem{Maldonado2026}
C. Maldonado, F. Nettel, and P.~A. S\'anchez,
\emph{Gravitational wave polarization modes and the kinematical tensors in
general relativity and beyond},
Eur. Phys. J. C \textbf{86}, 323 (2026),
\href{https://doi.org/10.1140/epjc/s10052-026-15567-6}{doi:10.1140/epjc/s10052-026-15567-6}.

\bibitem{AbebeMomeni2016}
A. Abebe, D. Momeni, and R. Myrzakulov,
\emph{Shear-free anisotropic cosmological models in $f(R)$ gravity},
Gen. Relativ. Gravit. \textbf{48}, 49 (2016),
\href{https://doi.org/10.1007/s10714-016-2046-1}{doi:10.1007/s10714-016-2046-1}.

\bibitem{SamiAbebe2020}
H. Sami and A. Abebe,
\emph{Perturbations of quasi-Newtonian universes in scalar-tensor gravity},
Int. J. Geom. Methods Mod. Phys. \textbf{18}, 2150158 (2021),
\href{https://doi.org/10.1142/S0219887821501589}{doi:10.1142/S0219887821501589}.

\bibitem{DeFelice2010}
A. De Felice and S. Tsujikawa,
\emph{$f(R)$ theories},
Living Rev. Relativ. \textbf{13}, 3 (2010),
\href{https://doi.org/10.12942/lrr-2010-3}{doi:10.12942/lrr-2010-3}.

\bibitem{Starobinsky1980}
A.~A. Starobinsky,
\emph{A new type of isotropic cosmological models without singularity},
Phys. Lett. B \textbf{91}, 99 - 102 (1980),
\href{https://doi.org/10.1016/0370-2693(80)90670-X}{doi:10.1016/0370-2693(80)90670-X}.

\bibitem{KodamaSasaki1984}
H. Kodama and M. Sasaki,
\emph{Cosmological perturbation theory},
Prog. Theor. Phys. Suppl. \textbf{78}, 1 - 166 (1984),
\href{https://doi.org/10.1143/PTPS.78.1}{doi:10.1143/PTPS.78.1}.

\bibitem{EllisVanElst1999}
G.~F.~R. Ellis and H. van Elst,
\emph{Cosmological models}, in \emph{Theoretical and Observational Cosmology},
edited by M. Lachi\`eze-Rey
(Kluwer, Dordrecht, 1999), pp. 1 - 116,
\href{https://arxiv.org/abs/gr-qc/9812046}{arXiv:gr-qc/9812046}.

\bibitem{Tsagas2008}
C.~G. Tsagas, A. Challinor, and R. Maartens,
\emph{Relativistic cosmology and large-scale structure},
Phys. Rep. \textbf{465}, 61 - 147 (2008),
\href{https://doi.org/10.1016/j.physrep.2008.03.003}{doi:10.1016/j.physrep.2008.03.003}.

\bibitem{Carloni2008}
S. Carloni, P.~K.~S. Dunsby, and A. Troisi,
\emph{The evolution of density perturbations in $f(R)$ gravity},
Phys. Rev. D \textbf{77}, 024024 (2008),
\href{https://doi.org/10.1103/PhysRevD.77.024024}{doi:10.1103/PhysRevD.77.024024}.

\bibitem{Molano2025}
D. Molano, F.~D. Villalba, L. Casta\~neda, and P. Bargue\~no,
\emph{Cosmological gauge invariant perturbation theory in $f(R)$ theories of
gravity}, Phys. Rev. D \textbf{111}, 024045 (2025),
\href{https://doi.org/10.1103/PhysRevD.111.024045}{doi:10.1103/PhysRevD.111.024045}.

\end{thebibliography}
\end{document}